\documentstyle{article}
\begin{document}
\title{Cohomology and Topological Anomalies}
\author{{\bf C.Ekstrand}
\\Department of Theoretical Physics, \\Royal Institute of
Technology, \\S-100 44 Stockholm, Sweden, \\ce@theophys.kth.se
}
\date{}
\maketitle
\newcommand{\eq}{\begin{equation}}
\newcommand{\eqend}{\end{equation}}
\newcommand{\eqa}{\begin{eqnarray}}
\newcommand{\eqaend}{\end{eqnarray}}
\newcommand{\nonu}{\nonumber \\ \nopagebreak}
\newcommand{\Ref}[1]{(\ref{#1})}
\newcommand{\B}{{\cal B}}
\newcommand{\F}{{\cal F}}
\newcommand{\cL}{{\cal L}}
\newcommand{\ep}{{\cal E}}
\newcommand{\W}{{\cal W}}
\newcommand{\M}{\cal M}
\newcommand{\A}{{\cal A}}
\newcommand{\T}{\cal T}
\newcommand{\D}{\cal D}
\newcommand{\G}{{\cal G}}
\newcommand{\Rr}{\cal R}
\newcommand{\Pp}{\cal P}
\newcommand{\C}{\cal C}
\newcommand{\V}{\cal V}
\newcommand{\N}{{\cal N}}
\newcommand{\U}{{\cal U}}
\newcommand{\nn}{\nonumber }
\def\eop{\nopagebreak\hfill $\Box$}
\newtheorem{definition}{Definition}
\newtheorem{lemma}{Lemma}
\newtheorem{theorem}{Theorem}
\newtheorem{proposition}{Proposition}
\newtheorem{corollary}{Corollary}
\begin{abstract}
The chiral anomaly can be considered as an object defined either on the space of gauge potentials or on the orbit space. We will discuss the relation between the two descriptions. We will also relate to the cohomology of the group of gauge transformations. 
\end{abstract}

\section{Introduction}
The chiral anomaly manifest itself in two  different guises. The first is as a lack of gauge invariance for the effective action. In this paper we will refer to this as the anomaly in the space-time formalism, or simply just the anomaly. The second is as an anomalous contribution to the equal-time commutator relations for the currents. The chiral anomaly is here referred to as the Schwinger term. Both the local anomaly and the Schwinger term can be described by forms either on the space of gauge potentials or on the orbit space. The relation between the two descriptions is called transgression. 

In this paper we will extend the transgression map to the global anomaly. It turns out that an extension is given by a map studied in \cite{FR}. This gives for instance a simple proof of the fact that the local anomaly always can be written as the transgression of a form. We will also discuss how the anomaly and Schwinger term can be described by the cohomology of the group of gauge transformations. 

In section 2 we define and discuss the anomaly and in section 3 we relate different descriptions of the anomaly in a commuting diagram. The classification of anomalies in terms of the cohomology of the group of gauge transformations is the content of section 4. In section 5 we extend parts of the construction to contain the Schwinger term. Finally, in section 6 we generalize to the case of a non-affine total space. 

 All manifolds and maps are assumed to be smooth. We will further assume that the orbit space $\A /\G$ admits a Leray (good) covering. The locality property (only polynomials in $A$, $dA$, $v$ and $dv$ are allowed) for the anomaly and Schwinger term will not be taken into account in this paper.

\section{The anomaly}
We will consider Weyl fermions coupled to an external gauge field $A\in\A$ in a $2n$-dimensional space-time $M$. The group $\G$ of gauge transformations consists of diffeomorphisms $g$ of a principal bundle $P\stackrel{G}{\rightarrow }M$  such that the base remains unchanged. It acts on the affine space $\A$ of connections on $P$ by pull-back: $A\cdot g :=g^\ast A$. To make the action free (so $\A /\G$ will be a smooth manifold) we will assume that $\G$ only consists of diffeomorphisms that leaves a fixed point $p_0\in P$ unchanged. 

The global anomaly is the lack of gauge invariance for the generating functional: 
\eq
\label{eq:1}
\exp (-W(A\cdot g))=-\exp (-W(A))f(A,g),
\eqend
where $W$ is the effective action. The global anomaly is thus given by a map $\A\times\G\rightarrow {\bf C^\times }$. Since the effective action is only defined up to (local) functionals in $\A$, $f(A,g)$ is defined up to a trivial factor $h(A\cdot g)h(A)^{-1}$ for some function $h:A\rightarrow {\bf C^\times }$. It follows directly from the definition that the consistency condition $f(A,g_1)f(A\cdot g_1,g_2)=f(A,g_1g_2)$ and the normalization condition $f(A,\mbox{id})=1$ has to be fulfilled. Together with the triviality condition this defines an element $[f]$ in a cohomology group $H^1(\G ,\mbox{Map}(\A ,{\bf C^\times }))$. That the group depends on the twist of the bundle $\pi :\A\rightarrow \A /\G$ is clear from the following statement \cite{K}:
\begin{proposition}
\label{p:1}
The global anomaly is zero if $\A\rightarrow \A /\G$ is trivial.
\end{proposition}
\vspace{2mm}
{\bf Proof}\hspace{2mm}
Since the bundle is trivial there exist a global section $s$. Define $g_s (A)$ by $A=s (\pi (A))g_s (A)$. By using the cocycle relation for $f$ it is then easy to see that $h(A)=f(s (\pi (A)),g_s (A))$ satisfies $f(A,g)=h(A\cdot g)h(A) ^{-1}$.
\eop

\noindent The group depends also on the topology of $\G$. For instance,
\begin{proposition}
\label{p:2}
The global anomaly is zero if $\G$ is simply connected.
\end{proposition}
\vspace{2mm}
{\bf Proof}\hspace{2mm}
From the exact sequence  
\[
0\rightarrow {\bf Z}\rightarrow {\bf C}  \stackrel{\mbox{\footnotesize{exp}} }{\rightarrow }  {\bf C^\times }\rightarrow 1\nonu 
\]
we obtain the exact sequence
\eqa
0 & \rightarrow & \mbox{Map}(\A\times\G ,{\bf Z})\rightarrow \mbox{Map}(\A\times\G ,{\bf C})\rightarrow \mbox{Map}(\A\times\G ,{\bf C^\times }) \nonu 
& \rightarrow & H^1(\A\times\G ,{\bf Z})\rightarrow ...\nn
\eqaend
Using the K\"unneth theorem and Hurewics theorem together with the fact that $\A$ is affine and $\G$ is simply connected gives $H^1(\A\times\G ,{\bf Z})=0$. The exact sequence then proves the existence of a $\theta \in \mbox{Map}(\A\times\G ,{\bf C}))$ such that $f(A,g)=\exp (\theta (A,g))$. Due to the normalization condition we can assume that $\theta (A,\mbox{id})=0$. The cocycle condition for $f$ implies that $\theta (A,g_1)+\theta (A\cdot g_1,g_2)-\theta (A,g_1g_2)=\chi (A,g_1,g_2)$, where $\chi: \A\times\G\times\G \rightarrow {2\pi i\bf Z}$. That $\theta $ is smooth implies that $\chi$ is smooth and therefore constant. The normalization condition then gives $\chi =0$ and $\theta $ therefore obeys a consistency condition. Let us now choose a covering $\U =\{ U_\alpha \}$ of $\A /\G$ such that $\A\rightarrow \A /\G$ is trivial over each $U_\alpha $. Then we choose corresponding sections $s_\alpha :U_\alpha \rightarrow \pi ^{-1}(U_\alpha )$ and a partition of unity $\{\rho _\alpha \}$ subordinate to the covering (it is known \cite{CMM} that any covering of $\A /\G$ has a refinement which admits a partition of unity). As in the proof of proposition \ref{p:1}, it is seen that $\xi (A)=\sum _\alpha \rho _\alpha \theta (s _\alpha (\pi (A)), g_{\alpha }(A))$ trivializes $\theta$. This implies that $h(A)=\exp (\xi (A)) $ trivializes $f$.
\eop

Let us now turn to the local anomaly. For this reason we insert $g=\exp (tX)$ in eq. \Ref{eq:1} and take the derivative at $t=0$:
\eqa
\partial _t\vert _{t=0}\exp (-W(A\cdot e^{tX})) & = & -\exp (-W(A))\partial _t\vert _{t=0}\exp (\log f(A,e^{tX}))\nonu
\Leftrightarrow
\partial _t\vert _{t=0}W(A\cdot e^{tX}) & = & \partial _t\vert _{t=0}\log f(A,e^{tX})\nn .
\eqaend
We thus see that the local anomaly 
\eq
\label{eq:2}
\omega (\A ,X)=\partial _t\vert _{t=0}\log f(A,e^{tX})\nn 
\eqend
 is equal to the gauge variation of the effective action. This also explains our choice of sign in front of the exponential in the right hand side of eq. \Ref{eq:1}. The consistency and triviality conditions carries over to define the local anomaly as an element in a cohomology group $H^1(\mbox{Lie}\G ,\mbox{Map}(\A ,{\bf C}))$. 

The local anomaly can be computed by the family index theorem. One then obtains a 2-form $\Omega$ on $\A /\G$, the curvature of the determinant line bundle. Lifted to $\A$ it can be written as:
\eq
\label{eq:3}
\pi ^{\ast }\Omega =c\int _M\mbox{tr}\left( (d+\delta )(A+v)+(A+v)^2\right) ^{n+1},
\eqend
where $c$ is a constant, $d$ is the exterior differential on $M$, $\delta$ is the exterior differential on $\A$ and $v$ is a 1-form on $\A$. We have assumed that $M$ is flat in order to omit the Dirac genus. It is well known that the local anomaly is the restriction to gauge (fibre) directions in $\A\rightarrow \A /\G$ of the connection of the determinant line bundle (as a bundle over $\A $). To obtain the local anomaly we must therefore write the form in eq. \Ref{eq:3} as $\delta$ on another form. Since $M$ is without boundary this can be accomplished by writing the integrand as $(d+\delta )$ acting on a form. It is well known that the Chern-Simon's form $\mbox{CS}(A,v)$ is such a form, i.e. 
\[
\pi ^{\ast }\Omega =\delta c\int _M\mbox{CS}(A,v).
\] 
The restriction of $c\int _M\mbox{CS}(A,v)$ to gauge directions is then the local anomaly. It is known that $\delta $ and $v$ becomes the BRST operator and the ghost in such directions.

The map from $\Omega $ to the local anomaly is called the transgression map. It is thus given by: pull-back to $\A$ and write $\pi ^{\ast }\Omega $ as $\delta $ of a form and restrict to gauge directions. Recall that locality plays an important role for transgression, but as we said in the introduction, this will be overlooked here. To give an explicit expression for the transgression map $T$ we introduce the map $\iota _A$ that imbeds $\G$ in a gauge orbit through $A\in\A$ by: $\iota _{A }(g)=A \cdot g$. Then we see that 
\eq
\label{eq:35}
T= \iota ^{\ast }_\bullet \delta ^{-1}\pi ^{\ast }
\eqend
 It is easy to check that $T$ is a well defined homomorphism $H_{\mbox{\footnotesize{dR}}}^{2}(\A /\G ,{\bf C})\rightarrow H^1(\mbox{Lie}\G ,\mbox{Map}(\A ,{\bf C}))$. 

Also the global anomaly can be related to the cohomology of $\A /\G$. The argument goes as follows \cite{FR}: Let $[f]\in H^1(\G ,\mbox{Map}(\A ,{\bf C^\times }))$ be given. A line bundle $\cL \stackrel{{ \mbox{\scriptsize\bf C}^{\times }}}{\rightarrow }\A /\G $ can then be defined through an equivalence relation $(A,\lambda )\sim (A\cdot g, f(A ,g)\lambda )$ in $\A \times {\bf C^\times }$. This defines a homomorphism from $H^1(\G ,\mbox{Map}(\A ,{\bf C^\times }))$ to the group of equivalence classes of line bundles over $\A /\G$. To define the inverse, let $p:\cL \stackrel{{ \mbox{\scriptsize\bf C}^{\times }}}{\rightarrow }\A /\G $ be given. By using that $\pi ^\ast \cL =\{ (A,l)\in\A\times \cL \vert \pi (A)=p(l)\}$ is trivial and thereby allows a global non-vanishing section it is easy to see how the inverse can be defined. Since the group of equivalence classes of ${\bf C^\times }$-bundles on $\A /\G$ is isomorphic to the \v{C}ech cohomology $\check{H}^1(\A /\G , \underline{{\bf C}}^\times )$ we have proven: $H^1(\G ,\mbox{Map}(\A ,{\bf C^\times }))\cong\check{H}^1(\A /\G , \underline{{\bf C}}^\times )$. In terms of representatives this map is given by $f\mapsto \{g_{\alpha\beta }\}$, where 
\eq
\label{eq:4}
g_{\alpha\beta }(\pi (A))=f(s _\alpha (\pi (A)),t_{\alpha \beta}(\pi (A))), 
\eqend
and the transition functions $t_{\alpha \beta}$ are defined by: $s_\beta =s_\alpha\cdot t_{\alpha \beta}$.

\section{A commuting diagram}
To obtain a commuting diagram we introduce the map from the equivalence class of a line bundle to the Chern class of its curvature. In terms of \v{C}ech cohomology this map is given by $[\{g_{\alpha\beta }\} ]\mapsto [\Omega ]\in H_{\mbox{\footnotesize{dR}}}^{2}(\A /\G ,{\bf C})$, where
\eq
\label{eq:5}
\Omega =\sum _{\alpha ,\beta }\rho _\beta \bar{\delta }\rho _\alpha \wedge \bar{\delta }\log g_{\alpha \beta }
\eqend
and $\bar{\delta }$ is the exterior differential on $\A /\G$. The kernel is the torsion (holonomy) and the image is the class of forms that satisfies the integrality condition. 
\begin{theorem}
\label{t:1}
The following diagram is commuting.
\eqa
&&\begin{array}{ccc}
 H^{1}(\G ,\mbox{Map}(\A,{\bf C^\times } )) & \longrightarrow & H^{1}(\mbox{Lie}\G ,\mbox{Map}(\A ,{\bf C}))\nonu
\downarrow \cong && \uparrow \nonu
\check{H}^1(\A /\G ,\underline{{\bf C}}^\times ) & \longrightarrow  & H_{\mbox{\footnotesize{dR}}}^{2}(\A /\G ,{\bf C})\nn 
\end{array}\nonu
\eqaend
The upper homomorphism is the map from the global anomaly to the local anomaly, eq. \Ref{eq:2}. The isomorphism to the left is given by eq. \Ref{eq:4}. The lower homomorphism is the map from the equivalence class of line bundles to the Chern class of their curvatures, eq. \Ref{eq:5}. The right map is the transgression homomorphism defined in eq. \Ref{eq:35}. 
\end{theorem}
\vspace{2mm}
{\bf Proof}\hspace{2mm}
Let $[f]\in H^{1}(\G ,\mbox{Map}(\A ,{\bf C^\times }))$ be given. 
It is mapped to $[\{ g_{\alpha\beta }\} ]\in $
$\check{H}^1(\A /\G ,\underline{{\bf C}}^\times )$ as in eq. \Ref{eq:4}.
 This maps to $[\Omega]\in H^2_{\mbox{\footnotesize{dR}}}(\A /\G ,{\bf C})$ through eq. \Ref{eq:5} according to:
\[
\Omega (\pi (A))=  \sum _{\alpha ,\beta}\rho _{\beta }(\pi (A))\bar{\delta }\rho _{\alpha }(\pi (A))\wedge \bar{\delta }\log f(s _\alpha (\pi (A)),t_{\alpha \beta}(\pi (A))).
\]
We now use $g_{s_\alpha }(A)=t_{\alpha \beta }(\pi (A))g_{s_\beta }(A)$ and the cocycle property of $f$ to obtain:
\[
\pi ^\ast \Omega (A) =  \sum _{\alpha}\delta\pi ^\ast  \rho _{\alpha}(A)\wedge \delta\log f(s _\alpha (\pi (A)),g_{s_\alpha }(A))
\]
This gives:
\eqa
(\iota ^{\ast }_\bullet \delta ^{-1}\pi ^{\ast }\Omega)(X)_A & = & \partial _{t}|_{t=0}\sum _\alpha \pi ^\ast \rho _{\alpha }(A)\log f(s _\alpha (\pi (A\cdot e^{tX})),g_{s _\alpha }(A\cdot e^{tX}))\nonu 
& = & \partial _{t}|_{t=0}\sum _\alpha \pi ^\ast \rho _{\alpha }(A)\log f(s _\alpha (\pi (A)),g_{s _\alpha }(A)\cdot e^{tX})\nonu 
& = & \partial _{t}|_{t=0}\sum _\alpha \pi ^\ast \rho _{\alpha }(A)\log f(A,e^{tX})\nonu 
& = & \partial _{t}|_{t=0}\log f(A,e^{tX})\nn ,
\eqaend
where the cocycle property for $f$ was used in the third equality. This agrees with eq. \Ref{eq:2} and the commutativity of the diagram has been proven.
\eop

The following statements follows directly:
\begin{corollary}
\label{c:1}
The local anomaly can always be written as the transgression of a 2-form (on the orbit space) that satisfies the integrality condition.
\end{corollary}

\section{The Serre spectral sequence}
We will now classify the local anomaly in terms of the cohomology of the group of gauge transformations. We will use spectral sequence techniques, whose basics can be found in most books on algebraic topology. Let $\U =\{ U_\alpha \} _{\alpha \in I}$ be a Leray covering of $\A /\G$. Consider the double complex associated with the bundle $\A\rightarrow\A /\G$: $E^{p,q}=\{ \Omega ^p(\pi ^{-1}(U_{\alpha _0}\cap ...\cap U_{\alpha _q} ),{\bf C})\} $, where $\Omega ^p(U)$ is the set of ${\bf C}$-valued $p$-forms on $U$. It is equipped with two differential operators induced by the \v{C}ech coboundary operator $\partial :E^{p,q}\rightarrow E^{p,q+1}$ and the exterior differential $\delta $ on $\A:E^{p,q}\rightarrow E^{p+1,q}$. Then $E^{p,q}_2 =H_\partial H_\delta E^{p,q}$ is such that $E_2^{0,n}\cong \check{H}^n(\A /\G ,{\bf C})$ and $E_2^{n,0}\cong H^n_{\mbox{\footnotesize{dR}}}(\G ,{\bf C})$. Interchanging the roles of the differential operators we get $E^{\prime \, p,q}_2 =H_\delta H_\partial E^{p,q}$ which is zero for $q\geq 1$ and $E^{\prime \, n,0}_2\cong H^n_{\mbox{\footnotesize{dR}}}(\A ,{\bf C})$ which is zero since $\A$ is affine. This implies that $E_{\infty }^{p,q}=0$. The exact sequence 
\[
0\rightarrow {E}_{3}^{1,0} \rightarrow  {E}_{2}^{1,0}\stackrel{{d}_2}{\rightarrow }{E}_2^{0,2}\rightarrow {E}_3^{0,2}
\rightarrow 0 
\]
together with the facts ${E}_{3}^{1,0} \cong {E}_{\infty }^{1,0} =0$ and ${E}_{3}^{0,2} \cong {E}_{\infty }^{0,2} =0$ implies then that:
\begin{proposition}
\label{p:3}
\[
d_2: H^1_{\mbox{\footnotesize{dR}}}(\G ,{\bf C})\rightarrow \check{H}^2(\A /\G ,{\bf C})
\]
is an isomorphism. 
\end{proposition}
 The statement is important for the anomaly. Indeed, we have shown that the local anomaly can be classified by the subgroup of $H^2_{\mbox{\footnotesize{dR}}}(\A /\G ,{\bf C})$ consisting of forms that satisfy the integrality condition. This subgroup is isomorphic to $\check{H}^2(\A /\G ,{\bf C})$ and proposition \ref{p:3} then states that the local anomaly can be classified by $H^1_{\mbox{\footnotesize{dR}}}(\G ,{\bf C})$.

In \cite{FR} a similar statement was proven. They showed that $H^1(\G ,{\bf Z})\cong {H}^2(\A /\G ,{\bf Z})$ under the assumption that $\G$ is connected, $\pi _0(\G )=1$. We will now show that this can be obtained as a corollary of proposition \ref{p:3}. Indeed, the exact homotopy sequence for a fibre bundle gives $\pi _{n+1}(\A /\G )\cong \pi _n(\G )$ so $\pi _1(\A /\G )=1$. In this case there is no holonomy so ${H}^2(\A /\G ,{\bf Z})\cong \check{H}^2(\A /\G ,{\bf C})$ and $H^1(\G ,{\bf Z})\cong H^1_{\mbox{\footnotesize{dR}}}(\G ,{\bf C})$ from which the statement follows. We thus see that the local and global anomaly are equal in this case. The statement can alternatively be proven by Hurewics theorem: $H^1(\G ,{\bf Z})\cong \pi _1(\G )\cong\pi _2(\A /\G )\cong {H}^2(\A /\G ,{\bf Z})$ (in fact, the above spectral sequence methods are often used to prove Hurewics theorem, \cite{BT}). In many interesting cases (when the space-time is a sphere of arbitrary dimension) the group $H^1(\G ,{\bf Z})\cong \pi _1(\G )$ is given by the homotopy of the gauge group. For instance, when $\G =\mbox{Map}(S^4,G)$ we have $\pi _1(\G )\cong \pi _5(G)$. Notice that when $\G$ is simply connected, $H^1(\G ,{\bf Z})=0$ and we have given an alternative proof of proposition \ref{p:2}.

\section{The Schwinger term}
We will now generalize the above results as far as possible so they include the Schwinger term as well. It is well known that the Schwinger term is classified by a group $H^2(\mbox{Lie}\G ,\mbox{Map}(\A ,{\bf C}))$ which is obtained from a generalization of $H^1(\mbox{Lie}\G ,\mbox{Map}(\A ,{\bf C}))$. The Schwinger term can be written as the transgression of a form in the subgroup of $H^3_{\mbox{\footnotesize{dR}}}(\A /\G ,{\bf C})$ consisting of forms that satisfies the integrality condition, see \cite{CE} for instance. This subgroup can be extended to $\check{H}^2(\A /\G ,\underline{{\bf C}}^\times )$ by adding torsion. In this way a global Schwinger term is obtained. Actually, also in this case there exists natural homomorphisms $H^2(\G ,\mbox{Map}(\A ,{\bf C}^\times ))\rightarrow H^2(\mbox{Lie}\G ,\mbox{Map}(\A ,{\bf C}))$, corresponding to the upper homomorphism, and $H^2(\G ,\mbox{Map}(\A ,{\bf C}^\times ))\rightarrow \check{H}^2(\A /\G ,\underline{{\bf C}}^\times )$, corresponding to the left homomorphism in the commuting diagram, see \cite{K} for the construction of the latter. However, due to two reasons we will not discuss these maps further: The first is that this does not lead to a commuting diagram. This can be seen from the fact that the composition of the left, the lower and the right homomorphism gives zero in this case. The second reason is that $H^2(\G ,\mbox{Map}(\A ,{\bf C}^\times ))$ is not the relevant group for the global Schwinger term. Indeed, proposition \ref{p:2} goes through in this case as well and states that $H^2(\G ,\mbox{Map}(\A ,{\bf C}^\times ))=0$ when $\G$ is simply connected. On the other hand it is known that the Schwinger term can be non-zero in this case. 

Let us now use the Serre spectral sequence to relate the classifying group $\check{H}^3(\A /\G ,{\bf C})$ for the Schwinger term to the cohomology of the group of gauge transformations. In this case we are interested in the isomorphism ${d}_3: {E}_3^{2,0}\rightarrow {E}_3^{0,3}$. Using ${E}_3^{0,3}={E}_2^{0,3}$ and 
\[
0\rightarrow {E}_3^{2,0}\rightarrow {E}_2^{2,0}\stackrel{{d}_2}{\rightarrow } {E}_2^{1,2}
\]
we see that it is ${E}_2^{1,2}$ that prevents us from getting an isomorphism between $H^2_{\mbox{\footnotesize{dR}}}(\G ,{\bf C})$ and $\check{H}^3(\A /\G ,{\bf C})$. 
\begin{proposition}
\label{p:4}
If $\G $ is simply connected, then
\[
{d}_3: H^2_{\mbox{\footnotesize{dR}}}(\G ,{\bf C})\rightarrow \check{H}^3(\A /\G ,{\bf C})
\]
is an isomorphism. 
\end{proposition}
\vspace{2mm}
{\bf Proof}\hspace{2mm}
Notice first that $\A /\G$ is connected since $\A$ is affine: lift two points in $\A /\G$ to $\A$ and then project down a path between the lifted points. Further, recall that $\pi _{n+1}(\A /\G )\cong \pi _n(\G )$. Since $\A /\G$ then is simply connected we obtain $\bar{E}_2^{1,2}\cong \check{H}^2(\A /\G ,{\bf C})\otimes H^1_{\mbox{\footnotesize{dR}}}(\G ,{\bf C})$, see for instance \cite{BT}. From the vanishing homotopies we now see that ${E}_2^{1,2}=0$ and the statement follows. 
\eop 

\noindent The assumption in the proposition is fulfilled for example when $\G$ is the group of maps from the circle to a compact and simply connected gauge group, \cite{PS}.

The fact that $\G$ is simply connected together with $\pi _{n+1}(\A /\G )\cong \pi _n(\G )$ implies that ${H}^3(\A /\G ,{\bf Z})\cong \check{H}^3(\A /\G ,{\bf C})$ and $H^2(\G ,{\bf Z})\cong H^2_{\mbox{\footnotesize{dR}}}(\G ,{\bf C})$. From proposition \ref{p:4} we then get $H^2(\G ,{\bf Z})\cong {H}^3(\A /\G ,{\bf Z})$. As for the anomaly this can also be proven with Hurewics theorem: $H^2(\G ,{\bf Z})\cong \pi _2(\G )\cong\pi _3(\A /\G )\cong {H}^3(\A /\G ,{\bf Z})$. Again, in the case when the physical space is a sphere of arbitrary dimension this group can be computed from the homotopy of the gauge group. For instance, when $\G =\mbox{Map}(S^3,G)$ we have $\pi _2(\G )\cong \pi _5(G)$, the same homotopy group as for the anomaly in the corresponding case.

It is known that the Schwinger term is related to extensions of $\G$. It is interesting that the assumption of a simply connected $\G$ plays an important role here as well, \cite{PS}.

\section{Generalization to a non-affine total space}
Anomalies can also arise in similar situations as above but with an $\A$ that is not affine. This can occur for instance in string theories and for $\sigma $-models. At first sight this general case seems to be problematic since the definition of transgression uses the fact that $\A$ has no cohomology. However, from the commuting diagram we see the forms in the image of the composition of the left and the lower homomorphism always can be transgressed. We have thereby proven that the commutative diagram holds for any manifold $\A$. Notice though that the left map in the commuting diagram is not necessary an isomorphism. Indeed, it is easy to see that
\[
0\rightarrow H^1(\G ,\mbox{Map}(\A ,{\bf C}^\times ))\rightarrow \check{H}^1(\A /\G ,\underline{{\bf C}}^\times )\rightarrow \check{H}^1(\A  ,\underline{{\bf C}}^\times )
\]
is an exact sequence. 

For the Serre sequence additional complications arise in this general case. The map $d_2$ in proposition \ref{p:3} is still defined. However, it is not an isomorphism. The map $d_3$, on the other hand, does in general not induce map from $H^2_{\mbox{\footnotesize{dR}}}(\G ,{\bf C})$ to $\check{H}^3(\A /\G ,{\bf C})$, even if $\G$ is simply connected. 

\thanks{\bf Acknowledgments:} This work was performed within the program \lq Quantization, generalized BRS cohomology and anomalies\rq $ $ at the Erwin Schr\"odinger Institute (ESI). I would like to thank Prof. R.A. Bertlmann and ESI for their kind hospitality.


\newpage
\begin{thebibliography}{}
\bibitem{FR} Falqui, G., Reina, C., Commun. Math. Phys. {\bf 102}, 503 (1985)
\bibitem{K} Kelnhofer, G., J. Math. Phys. {\bf 33}, 2071 (1992)
\bibitem{CMM} Carey, A. L., Mickelsson, J., Murray, M. K., 
     Comm. Math. Phys. {\bf 183}, 707 (1997)
\bibitem{BT} Bott, R., Tu, L. W., 
{\sl Differential Forms in Algebraic Topology}, Springer-Verlag,
Berlin Heidelberg New York, 1982
\bibitem{CE} Ekstrand, C., hep-th/0002063
\bibitem{PS} Pressley, A., Segal, G., Loop Groups, Clarendon Press, Oxford, 
1986 
\end{thebibliography}
\end{document}